\documentclass[12pt,preprint]{aastex}

\newcommand{\chisq}{\mbox{$\chi^2$ }}	    
\newcommand{\degrees}{\mbox{$^{\circ}$}}    

\newcommand{\NH}{\mbox{${\rm N}_{\rm H}$}} 

\newcommand{\meszaros}{M\'{e}sz\'{a}ros}
\newcommand{\peterm}{M\'{e}sz\'{a}ros}

\newcommand{\nature}{{\nat}}
\newcommand{\etal}{{et al. }}		

\newcommand{\chandra}{\mbox{\it Chandra}}	    

\newcommand{\swift}{{\it Swift}}
\shorttitle{{\it Chandra} Observations of GRB\,051221A}
\shortauthors{Burrows, D. N., et al.}

\begin{document}


\title{Jet Breaks in Short Gamma-Ray Bursts.  II: The Collimated
  Afterglow of GRB\,051221A}


\author{
David N. Burrows\altaffilmark{1}, 
Dirk Grupe\altaffilmark{1},
Milvia Capalbi\altaffilmark{2},
Alin Panaitescu\altaffilmark{3},
Sandeep K. Patel\altaffilmark{4,5},
Chryssa Kouveliotou\altaffilmark{5},
Bing Zhang\altaffilmark{6},
Peter \meszaros\altaffilmark{1,7},
Guido Chincarini\altaffilmark{8,9},
Neil Gehrels\altaffilmark{10}, 
Ralph A.M. Wijers\altaffilmark{11}
}

\altaffiltext{1}{Department of Astronomy and Astrophysics, Pennsylvania State
University, 525 Davey Lab, University Park, PA 16802; {\it
  burrows@astro.psu.edu}}
\altaffiltext{2}{ASI Science Data Center, via Galileo Galilei, 00044 Frascati, Italy}
\altaffiltext{3}{Space Science and Applications, MS D466, Los Alamos
  National Laboratory, Los Alamos, NM 87545}
\altaffiltext{4}{Universities Space Research Association, 10211
  Wincopin Circle, Suite 500, Columbia, MD 21044-3432}
\altaffiltext{5}{NASA/Marshall Space Flight Center, National Space Science
Technology Center, VP-62, 320 Sparkman Dr., Huntsville, AL 35805}
\altaffiltext{6}{Department of Physics, University of Nevada, Las Vegas, 
NV 89154}
\altaffiltext{7}{Department of Physics, Pennsylvania State University,
University Park, PA 16802}
\altaffiltext{8}{INAF -- Osservatorio Astronomico di Brera, Via
  Bianchi 46, 23807 Merate, Italy}
\altaffiltext{9}{Universit\`a degli studi di Milano-Bicocca,
                 Dipartimento di Fisica, Piazza delle Scienze 3, I-20126 Milan, Italy}
\altaffiltext{10}{NASA Goddard Space Flight Center, Greenbelt, MD 20771}
\altaffiltext{11}{Astronomical Institute 'Anton Pannekoek', University of
Amsterdam, Kruislaan 403, NL-1098 SJ Amsterdam, The Netherlands}




\begin{abstract}
We report the best evidence to date of a jet break in a short
Gamma-Ray Burst (GRB) afterglow, using 
\chandra\ and \swift\ XRT observations of the X-ray
afterglow of GRB\,051221A.  The combined X-ray
light curve, which has three breaks, is similar to those
commonly observed in \swift\ observations of long GRBs.
A flat segment of the light curve at $\sim 0.1$ days after the
burst represents the first
clear case of strong energy injection in the external shock of a short
GRB afterglow.
The last break in the light curve occurs at $\sim 4$~days
post-burst and breaks to a power-law decay index of $\sim 2$.
We interpret this as a jet break, with important
implications for models of short GRBs, since it requires collimation of
the afterglow into a jet with an initial opening angle 
$\theta_0 \sim 4\degrees-8 \degrees$ and implies a
total jet kinetic energy of $E_{jet} \sim (1-5)\times 10^{49}$ erg.
Combined with the lack of a jet break in GRB\,050724, this suggests a
wide range in jet collimation in short GRBs, with at least some having
collimation similar to that found in long GRBs, though with significantly
lower jet energies.
\end{abstract}


\section{Introduction}

Until recently, afterglows of short Gamma-Ray Bursts  have
proven to be extremely elusive, frustrating efforts to identify their
progenitors and environments.  This situation changed dramatically in
2005.  The \swift\ \citep{Gehrels04} Burst Alert Telescope
\citep[BAT;][]{Barthelmy05a} localized 11 short Gamma-Ray Bursts (GRBs)
between 2005 February 1 and 2006 January 31, and the \swift\ 
X-Ray Telescope \citep[XRT;][]{Burrows05a} identified X-ray afterglows
from six of these (two could not be observed by the XRT,
and the rest were too faint to detect).  These observations led to the
first precise localization of a short GRB, GRB\,050509B, and
the discovery that it was probably located in a giant elliptical
galaxy with extremely low star formation \citep{Gehrels05,Hjorth05b,Bloom06}, lending support to a merger
model for short bursts
\citep[e.g.][]{Lattimer76,Paczynski86,Eichler89,Paczynski91,Rosswog03,Rosswog05}.  
This picture was confirmed two months later
by the {\it HETE-II} discovery of the short GRB\,050709 \citep{Villasenor05}
and its localization to a region of low star formation \citep{Hjorth05a,Fox05};
and by the \swift\ discovery of the short GRB\,050724 \citep{Barthelmy05b,Campana06} in
another elliptical galaxy.

Of the six short bursts localized by the XRT, only three were bright
enough in X-rays to permit detailed study: GRB\,050724, GRB\,051221A, and
GRB\,051227 \citep{Barbier06}.  
\chandra\ followup observations obtained for GRB\,050724 and for
GRB\,051221A are critical for constraining jet breaks in both bursts.
GRB\,050724 is discussed in a companion paper \citep[][Paper I]{Grupe06}.  In
this paper we discuss the late X-ray afterglow of GRB\,051221A as
observed by the XRT and by the \chandra\ ACIS-S instrument.  
We show that a break is observed in the late-time X-ray light curve.  If interpreted as
a jet break, these observations provide a measurement of the jet opening angle, 
and hence allow us to determine the jet energy for this short burst.

We characterize
the dependence of the X-ray flux on time and frequency as $F(t,\nu)
\propto (t-T_0)^{-\alpha} \nu^{-\beta}$, where $T_0$ is the time of
the BAT trigger, $\alpha$ is the decay index, and $\beta$ is the
spectral energy index.  Error bars on data points are $1 \sigma$,
while those given for model parameters are 90\%
confidence limits for one interesting parameter ($\chi^2_{min}+2.7$) unless otherwise
specified.  We use standard $\Lambda$CDM cosmological parameters of 
$\Omega_{\rm M}$=0.27, $\Omega_{\Lambda}$=0.73 and $H_0$=71 km
s$^{-1}$ Mpc$^{-1}$.

\section{\label{observe} Observations and Data Reduction}
GRB\,051221A was detected by the \swift/BAT at ${\rm T}_0$=01:51:16~UT on 21
December 2005 \citep{Parsons05}.  
The burst was a short,
hard burst, with $T_{90} = 1.4 \pm 0.2$~s and a hard
photon index in the 15-150 keV band of $\Gamma = 1.39 \pm 0.06$
\citep{Cummings05}.
The Konus-Wind instrument measured a cut-off power law spectrum with 
$E_{peak}=402^{+93}_{-72}$~keV \citep{Golenetskii05}.  
The 15--150~keV fluence was $1.16 \pm
0.04 \times 10^{-6}$ ergs cm$^{-2}$ \citep{Cummings05}, making this the most fluent
BAT-detected short burst yet, by a factor of two.

The \swift/XRT observations of GRB\,051221A began at 01:52:44~UT, 88 seconds after
the BAT trigger.  The XRT was not able to determine a
centroid on-board because of insufficient counts in the 2.5 second
Image Mode exposure \citep[see][for a description of the XRT readout
modes]{Hill04}.  
Data were taken primarily in Windowed Timing mode
from $\rm T_0+93$~s until about $\rm T_0+300$~s, after which the instrument switched to
Photon-Counting (PC) mode.
Subsequent ground analysis of the PC mode data provided a 
position with $3\farcs5$ uncertainty \citep{Burrows05b}.  
XRT observations continued until 2006
January 3 (13.7~days after the trigger), although the source
became undetectable by XRT after about 11 days post-burst, and only upper
limits are available at later times.

The XRT data were reduced with the XRTDAS tools included in the HEAsoft 6.0.4 package,
using the latest calibration files available in CALDB and applying
standard data screening.
For the WT mode data, events  
in the 0.3--10 keV band with grades 0--2 were used in the analysis
\citep[see][for XRT event grade definitions]{Burrows05a}. 
For the PC mode data, we selected events in the same energy range with 
grades 0--12.

The GRB afterglow was located near a hot column on the CCD.  The
position on the detector changes with each orbit, leading to a
significant loss of effective area for some orbits during which the
PSF core falls partially on the columns that are masked
off by the analysis software.  In order to minimize
the impact on our data analysis, we used a small data extraction
region (a 5 pixel radius circle in PC mode and a 10 pixel box width in
WT mode), correcting the derived count rates for the fraction of the
PSF included in the extraction regions. (The plate scale is
$2\farcs34$ per pixel.)  When the hot columns
penetrated the extraction region we used a larger extraction area (10
pixel radius circle) and calculated exposure maps, excluding
the hot columns and adjacent columns, to make the necessary
corrections to the count rates.
These corrections ranged up to a factor of 1.9 for some orbits. 

Data in the first 200~s of PC mode, when the source count rate exceeds
$\sim 0.5$~counts~s$^{-1}$, were affected by pile-up, and were
corrected by excluding events within a circle of 2 pixels radius
around the afterglow \citep[see][for details of
pile-up correction procedures for XRT data]{Vaughan06,Pagani06,Romano06}.  
Again, these effective area losses were taken
into account in calculating the light curve.  
The background used in each mode for
both light curve and spectral analysis was estimated from nearby
source-free regions.

\bigskip
The \chandra\ Observatory performed five Target of Opportunity
observations with the ACIS-S3 CCD under our AO7 observing program, spanning the
interval from ${\rm T}_0+1.2 \times 10^5$~s to ${\rm T}_0+2.3 \times 10^6$~s.  The observing
log is given in Table~\ref{tbl:obs_log}.  The first two \chandra\ observations overlap
the \swift/XRT data and provide valuable spectral information at late
times, when the count rate in the XRT is too low for spectroscopy.
The last three \chandra\ data points extend the X-ray light curve of
GRB\,051221A out to 16 January 2006.  The observations were performed
in Faint or Very Faint mode with the standard 3.2~s frame time.  Data were
reduced using version 3.3 of the CIAO software with CALDB version
3.2.1.  
Events from the GRB afterglow were selected using a source extraction
radius of $R=1 \farcs 75$ for the 2005 December
observations and $R=1 \farcs 0$ for the 2006 January observations.
Background regions were chosen from a source-free area using a radius
10 times larger than the source region.

\section{\label{results} Data Analysis}

\subsection{Position}
\begin{figure*}
\centering
\includegraphics[width=4in]{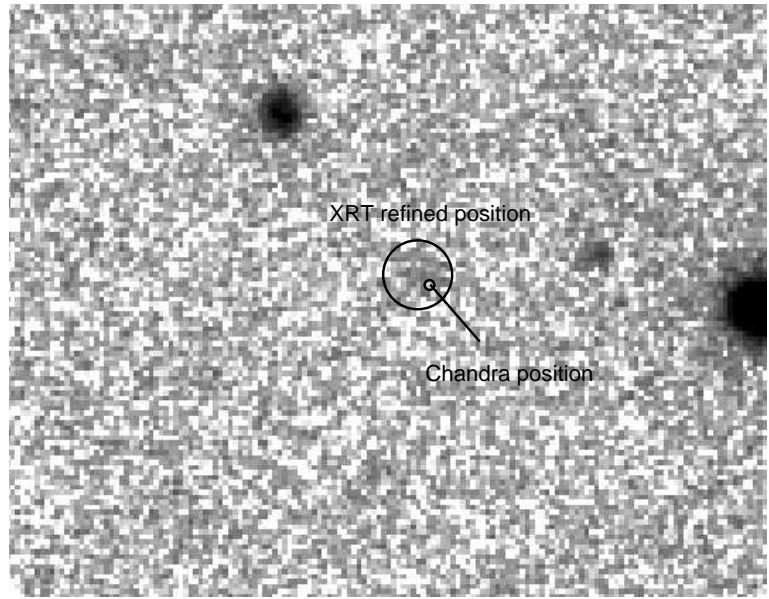}
\caption{XRT and \chandra\ error circles plotted over the
  UVOT/UVW1-band image
  (narrow band UV filter centered at 251~nm; 3.3~ks integration) of the GRB field. }
\label{fig:image}
\end{figure*}

A preliminary \chandra\ position for the afterglow was given in
\citet{Grupe05}, based on the standard \chandra\ attitude solution,
which is typically good to about $0\farcs5$.
We have improved the astrometry of the first \chandra\ observation by
reference to the 2MASS system.  The \chandra/ACIS data
were reprocessed with pixel randomization turned off in order to
provide the most accurate positions.  We find 33 X-ray sources on CCDs
S2 and S3 within 6 arcminutes of the GRB position with signal-to-noise
ratio greater than 2.7.  These were matched to the 2MASS
catalog to look for near-IR counterparts, resulting in 6 potential
matches.  Two of these were found to have large offsets and were
discarded as unrelated position coincidences.  We were left with 4
optical counterparts to X-ray sources, all with offsets less than
1$\sigma_{PSF}$.
Averaging the offsets between the \chandra\ and 2MASS positions for
these four objects, we find astrometry corrections of $-0\farcs094$
in right ascension and $+0\farcs015$ in declination.
We applied these offsets to the best-fit \chandra\ position (using a
circular Gaussian fit to the image) to obtain the final position of
the X-ray afterglow: 
\begin{displaymath}
(\alpha,\delta)_{J2000}=(21^{\rm h} 54^{\rm m} 48\fs620,
+16\degrees 53' 27\farcs19).
\end{displaymath}
The RMS residuals are $0\farcs17$ in RA and $0\farcs25$ in
declination.
This position is $0\farcs28$ arcseconds from the optical afterglow
\citep{Soderberg06}, which was calibrated relative to the USNO-B
system and has $0\farcs18$ rms uncertainty in each coordinate.  
We note that the optical observations are
better-suited to determining the offset of the GRB from its host
galaxy, which we do not detect with \chandra; this offset is given by
\citet{Soderberg06} as $0\farcs12 \pm 0\farcs04$.
The XRT and \chandra\ positions are shown overlaid on a UVOT UVW1 band
image in Figure~\ref{fig:image}.

\subsection{\label{sec:spectral}Spectral Analysis}

XRT spectra were accumulated from the WT mode data and from the PC
mode data.  
In the latter case, only data from $\rm T_0+501$~s to $\rm T_0+65721$~s were used, as
the earlier PC mode data suffered from pile-up, which distorts
spectral fits.
Data were binned to have at
least 20 counts per bin and were fitted to an absorbed power law, using {\it XSPEC} version
12.2.1 with version 007 (20060104) of the XRT response matrices.  
Results of the spectral fits are given in Table~\ref{tbl:spectral}.
From the PC mode spectrum,
which is averaged over the first three segments of the light curve, 
we find an excellent fit to an absorbed power
law with photon index $\Gamma = 2.1 \pm 0.2$ and $\NH =
(1.8^{+0.7}_{-0.5}) \times 10^{21}$~cm$^{-2}$ ($\chi^2 = 13.5$ for 30 degrees of freedom).  
These results are in good agreement with those from the WT mode
spectrum, which has fewer counts and larger uncertainties.
The absorbing column density is $\sim 2\sigma$ higher than the Galactic
value of $6.7 \times 10^{20}$~cm$^{-2}$.

The \chandra\ spectra
were analyzed using {\it XSPEC} version 12.2.1, with data binned to have
at least 20 photons per energy bin.
Spectral fit results for absorbed power laws are given in Table~\ref{tbl:spectral} for the
first two \chandra\ observations individually and for their combined spectrum.
For the combined spectrum, 
we obtained an excellent fit with photon index $\Gamma = 1.94^{+0.29}_{-0.19}$ and $\NH =
(1.4 \pm 0.9) \times 10^{21}$~cm$^{-2}$ ($\chi^2=17.2$ for 22 degrees
of freedom). 
These results are in good agreement with the \swift/XRT spectral
results, indicating no change in spectral parameters from the early
portion of the light curve to the late portion at $\sim 10^5$~s post-burst.

\subsection{\label{sec:lc}Light curve}
\begin{figure*}
  \includegraphics[angle=270,width=\textwidth, bb= 104 73 543  663, clip]{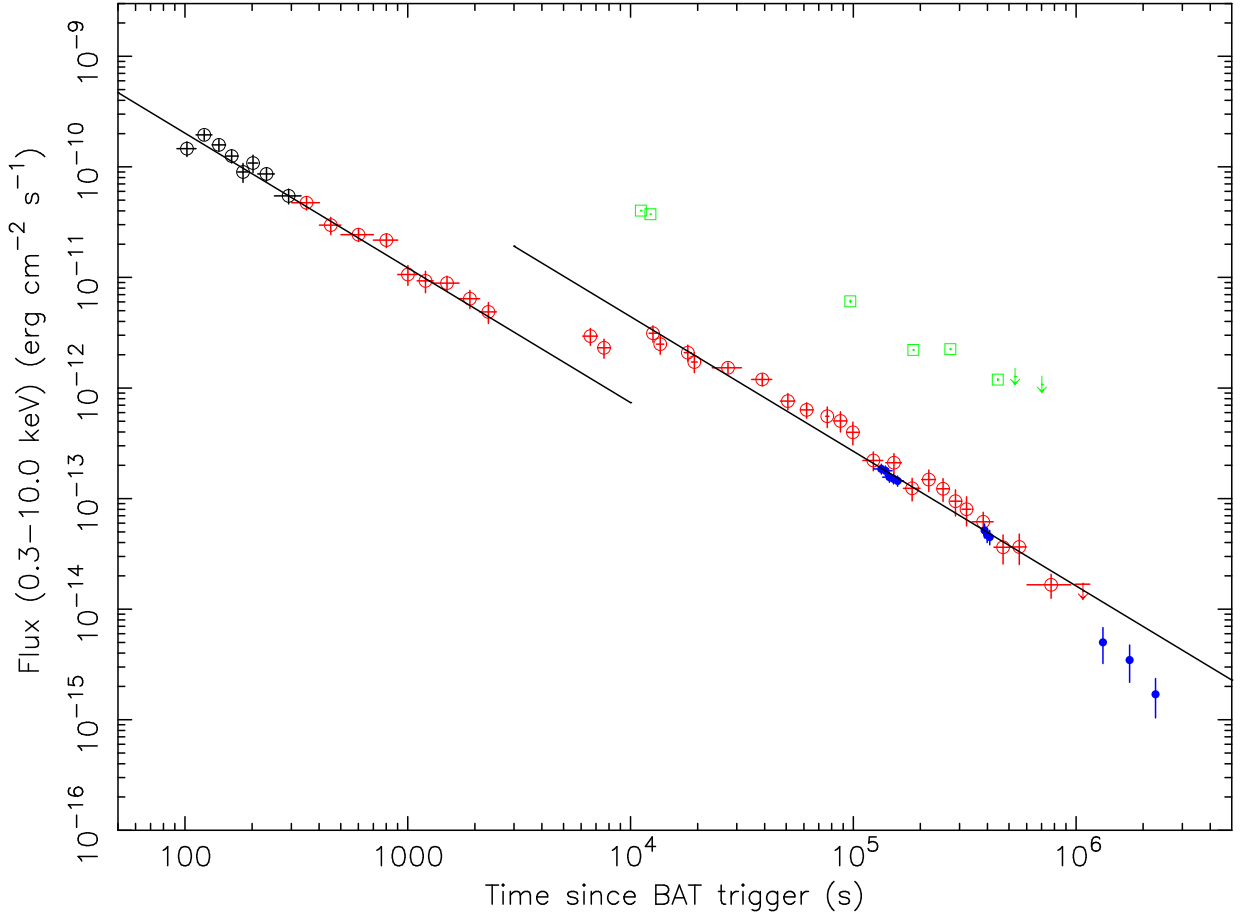}
\caption{\label{fig:lc} Combined \swift~XRT and \chandra~ACIS-S light
curve of the afterglow of GRB\,051221A. The light curve shows the 0.3-10.0
keV unabsorbed flux. Black circles indicate the WT mode XRT data
at early times.  Red circles indicate the PC mode XRT data.  Blue
dots indicate the \chandra\ observations at late times.  Green squares
(arrows) indicate the $r'$ band measurements (upper limits) 
from \citet{Soderberg06}, arbitrarily scaled
for comparison with the X-ray data.  The black lines have slopes of
-1.20, determined from a three segment fit to the data up to
$T_0+360$~ks with the first and third decay indices tied together (see
\S\ref{sec:lc}).
The X-ray light curve deviates from the late power law 
after the last $r'$-band detection.
}
\end{figure*}

Using the best fit spectral parameters for the XRT data, we obtained a
mean exposure-corrected
conversion factor between XRT count rate and unabsorbed source flux in
the 0.3--10 keV band of $7.9~\times~10^{-11}~({\rm erg~cm}^{-2}~{\rm s}^{-1}) / ({\rm count~s}^{-1})$
for this afterglow, taking into account the fraction of the PSF
contained within the source extraction region.
The XRT light curve shown in Figure~\ref{fig:lc} was generated by
multiplying the background-subtracted XRT count rate by the energy
conversion factor (ECF) appropriate for each orbit of data (since the
exposure varies as described in \S\ref{observe}).
A similar procedure was used to convert the \chandra\ count rates to
flux units, using the \chandra\ spectral fit results discussed in \S\ref{sec:spectral}
to calculate the \chandra/ACIS-S ECF for this
source, $1.07 \times 10^{-11}~{\rm erg~cm}^{-2}~{\rm count}^{-1}$.
Table~\ref{tbl:lc} gives the fluxes and uncertainties for all the
X-ray data points from Figure~\ref{fig:lc}.
For times periods before about 100 ks, our XRT fluxes are
systematically about 40\% higher than those reported by
\citet{Soderberg06}, with better agreement at later times.

As shown in Figure~\ref{fig:lc}, the XRT light curve can be
approximated as two power law segments of roughly equal slope separated by a
flatter segment.
The best-fit power law slope for the data from ${\rm T}_0+100$\,s to
${\rm T}_0+2200$\,s is $1.09^{+0.15}_{-0.10}$, 
while the third segment (from $3 \times 10^4 - 3.6 \times
10^5$\,s) has a best-fit decay index of $1.19 \pm 0.06$. 
%
The afterglow becomes undetectable in the XRT after 11 days, but the
\chandra\ observations at later times show clear evidence for a
deviation from the earlier power-law slope.

 Since the slopes of the first and third segments are consistent within the
uncertainties, we fixed these slopes to be the same in our following
analysis.  
We then obtained the following best-fit results to the entire X-ray
light curve:
\begin{itemize}
\item The initial decay index is $\alpha_1 = 1.20^{+0.05}_{-0.06}$,
\item the first break occurs at $t_{b1} = 3.70^{+0.54}_{-1.00}$~ks,
after which the decay index is $\alpha_2 = 0.04^{+0.27}_{-0.21}$, 
\item the second break occurs at $t_{b2} = 14.9^{+5.9}_{-2.8}$~ks 
and is followed by a decay index of 
$\alpha_3 \equiv \alpha_1$,
\item the third break occurs at $t_{b3} = 354^{+432}_{-103}$~ks and
is followed by a decay index of $\alpha_4 = 1.92^{+0.52}_{-0.25}$.
(The uncertainties in the final break time and final slope were
obtained by fixing $\alpha_3=1.20$ and calculating the confidence region in
the $t_{b3}$/$\alpha_4$ plane satisfying $\chi^2_{min}+4.6$, representing a
90\% confidence region for 2 interesting parameters.)
\end{itemize}
The \chandra\ data were critically important in determining the
parameters of the late portion of the light curve.  The fit to the third segment
is dominated by the small uncertainties in the first two \chandra\
observations, which were each binned into a
single point for this fit, although we show them
binned at finer time resolution in Figure~\ref{fig:lc}.
The last segment of the light curve depends almost entirely on the
\chandra\ data for definition of the break time and slope.
The overall fit has $\chi^2=21.4$ for 26 degrees of freedom.
By comparison, the best fit model with a single slope after 15 ks (no jet break) has 
$\alpha_3 = 1.39^{+0.04}_{-0.03}$ with $\chisq=59.9$ for 26 degrees of freedom.
Finally a fit with a single slope of 1.20 after 15 ks (as shown in Figure~\ref{fig:lc}) has 
$\chisq=135$ for 27 degrees of freedom.
We conclude that the third break in the light curve is required by the
data.

We note that the X-ray light curve shows evidence for 
a strong energy injection episode in the time range (2200--30,000s),
as discussed in more detail by \citet{Soderberg06}.
This is the first clear evidence for energy injection into the
external shock for a short GRB, and requires a mechanism for
refreshing the external shock similar to those proposed for long GRBs,
but operating in the context of short GRB models.

\section{\label{discuss} Discussion}
We interpret the last break in the X-ray light curve as a jet
break.  These data, based on a well-sampled and
well-behaved X-ray light curve, constitute the best measurement yet
of a jet break for a short GRB.    
The jet parameters can be
obtained by fitting a model of jet evolution to the available radio,
optical, and X-ray lightcurves.  Following \citet{Panaitescu03} and \citet{Panaitescu05},
we model the afterglow of GRB\,051221A with a uniform
relativistic jet undergoing 
lateral expansion and interacting with a homogeneous circumburst medium. The calculation
of the synchrotron emission from the electrons accelerated by the forward shock is done
assuming that the distribution of the electrons with energy is a power-law and that the
electron and magnetic field energies are constant fractions
($\epsilon_e$ and $\epsilon_B$, respectively)
of the post-shock energy.
Radiative losses, the spread in the photon arrival time due to the curvature of the emitting
surface, and the relativistic beaming of the radiation are taken into account. The emission
from the reverse shock sweeping into the out-flowing ejecta, synchrotron self-absorption (at radio
frequencies), the effect of inverse-Compton scattering on electron
cooling and its contribution 
to the X-ray light curve are also included but, for the best-fit parameters obtained
for 051221A, are not important.

The best-fit parameters are determined by minimization of $\chi^2$ between the model fluxes
and the X-ray, optical, and radio measurements. 
The basic jet model has six parameters: three for the jet dynamics (initial
energy $E_0$, initial opening $\theta_0$, and circumburst medium density $n$) and three for the
jet emission (the two microphysical parameters $\epsilon_e$ and $\epsilon_B$, and the slope 
$-p$ of the electron distribution). The available radio, optical, and X-ray measurements 
determine three or four characteristics of the afterglow synchrotron emission (peak flux, 
location of the injection and cooling break frequencies, and possibly
the self-absorption frequency), depending on whether the self-absorption 
frequency is below or above the radio
, which constrain $E_0$, $n$, $\epsilon_e$ and $\epsilon_B$. 
The spectral slope of the optical and X-ray
continua, as well as the optical and X-ray decay indices, over-constrain the electron index $p$.
The epoch of the last break in the X-ray light curve determines
primarily the initial jet opening angle.
Finally, the injected kinetic energy is constrained by the flattening
of the X-ray light-curve in its second segment.

\begin{figure*}
\includegraphics{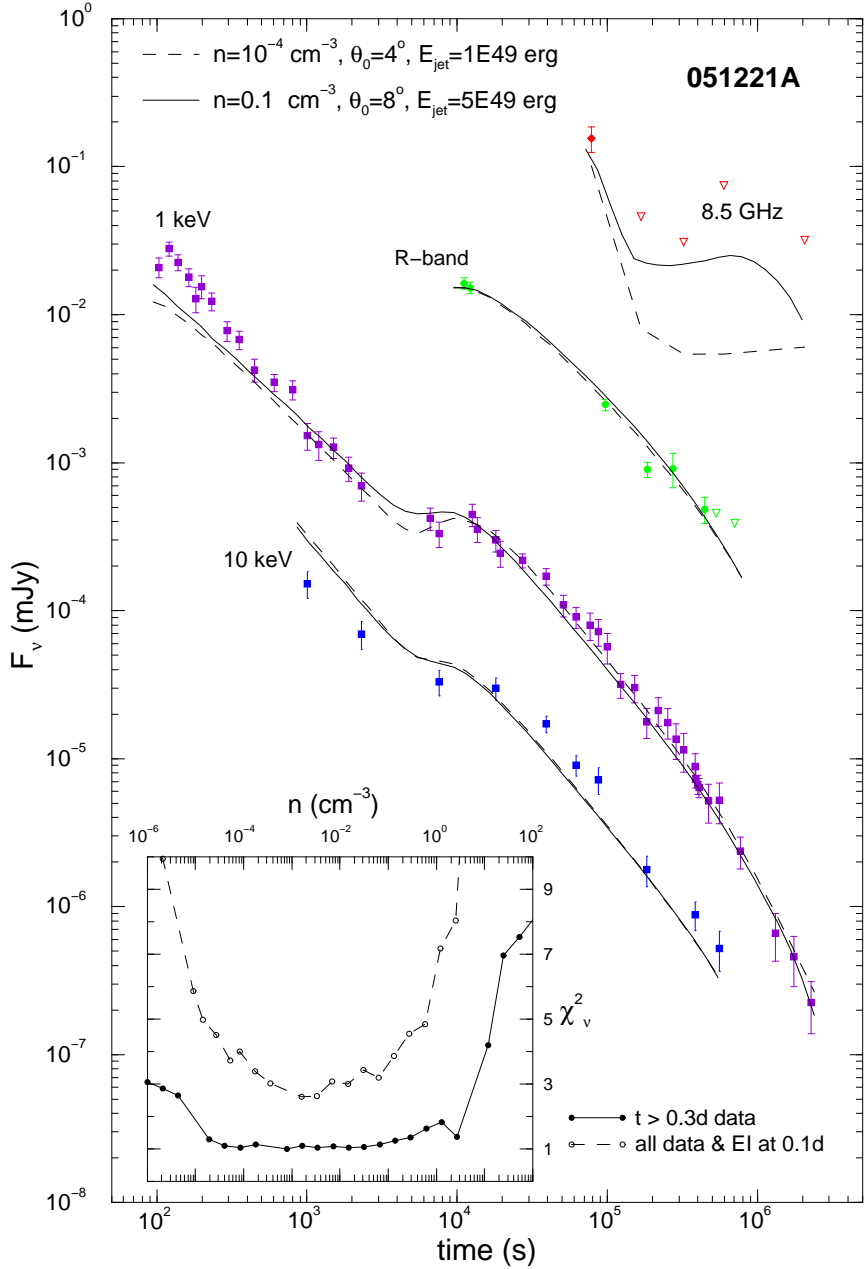}
\caption{ Examples of fits to the multiwavelength measurements of GRB afterglow 051221A
     obtained with the jet model and an episode of energy injection in
     the blast wave.
     The low density fit ($n=10^{-4}$~cm$^{-3}$) is shown with dashed
     curves, the higher density fit ($n=0.1$~cm$^{-3}$) with continuous
     lines. The radio and optical data points are taken
     from \citet{Soderberg06}. Triangles indicate $2\sigma$ radio
     upper limits.  (The fits also included $i'$ band and $z'$ band data, which
     are omitted from this figure for clarity).  
     Inset: variation of $\chi^2_\nu$ with circumburst
     medium density (in protons cm$^{-3}$) for the model fits. The
     dashed line is for a fit to all of the data, including an energy
     injection episode commencing at about 4000 s.  Better fits are obtained with 
     the jet model without energy injection if only the data after the X-ray flattening 
     are fitted (solid line). Similar jet parameters are obtained in
     both cases. }
\label{fig:fits}
\end{figure*}

Figure \ref{fig:fits} shows two fits to the radio, optical, and X-ray measurements of the
afterglow of GRB 051221A obtained with a relativistic jet interacting with a homogeneous
circumburst medium and undergoing energy injection at 0.1 days (to accommodate the 
flattening of the X-ray light-curves at that epoch). The two fits correspond to a lower limit
($n= 10^{-4}\, {\rm cm^{-3}}$) and an upper limit ($n=0.1\,
{\rm cm^{-3}}$) on the density of the ambient medium.
These densities span the allowed range, as shown in the inset by the variation of the
reduced $\chi^2_\nu$ with ambient density (the remaining model parameters were
left free to minimize $\chi^2$). For $n=10^{-4}$~cm$^{-3}$, which is characteristic for 
a binary merger occurring in the halo of the host galaxy, the jet initial opening
is $\theta_0 = 4\degrees$ and the total jet kinetic energy (after injection) is 
$E_{jet} = 10^{49}$ erg. For $n=0.1$~cm$^{-3}$, more typical of the
interstellar medium, the jet parameters are $\theta_0 = 8\degrees$ and $E_{jet} = 5\times
10^{49}$ erg. For either case, at the epoch of energy injection ($\sim 0.1$ d), the
incoming ejecta increase the forward-shock energy by a factor of 2. Also for either case,
the jet deceleration and decrease of the relativistic beaming of its emission renders
the boundary of the spreading jet visible to the observer at a few days, when the X-ray 
light-curve decay exhibits a steepening.
Given the small angular offset from the center of the host galaxy and
the evidence for a slight excess in \NH\ above the Galactic value, we
favor the higher density regime.

The fits obtained using the complete X-ray light curves 
are not statistically acceptable ($\chi^2_\nu \geq 2.5$ for 43 degrees 
of freedom). This is largely due to the poor fit at the earliest
times, when the decay rate of the pre-injection X-ray emission, 
$F_x \propto t^{-1.20}$, is faster than the model
prediction. The post-break decay of the 
X-ray emission, $F_x (t>4d) \propto t^{-1.93}$, indicates that the power-law
distribution with energy of the shock-accelerated electrons has an exponent $p=\alpha_x\simeq 2$. 
Together with the average spectral energy index of the X-ray continuum, $\beta_x
\equiv \Gamma-1 = 0.96\pm0.09$,
this implies that the cooling frequency is below the X-ray band (in which case $\beta_x = p/2$).
Then, the pre-injection decay of the X-ray light-curve should be $F_x \propto t^{-(3p-2)/4}
 = t^{-0.94\pm0.14}$, which is slower than observed, suggesting a departure from the
standard assumptions of the jet model (e.g. a slightly evolving index $p$ or non-uniform
ambient medium density). (We refer the reader to \citet{Zhang04} for a
compilation of the relations between $\alpha_x, \beta_x$, and $p$ for a
variety of models, together with references to the original sources of
these relationships.)

Statistically acceptable fits ($\chi^2_\nu \approx 1.0$)
are obtained if only the post-injection data are fit with the jet model, as shown
in the inset of Figure \ref{fig:fits}. The resulting jet parameters are very similar to 
those obtained for the energy injection model except that the range of allowed densities 
extends to larger values, $n \leq 3\; {\rm cm^{-3}}$. 

The single radio flux measurement may be dominated by the
reverse shock component, which depends on the
Lorentz factor of the ejecta responsible for the energy injection
episode.  With only a single radio detection, the Lorentz factor of the incoming ejecta 
is not well-constrained by the data (although the injected energy can
be constrained by the X-ray data).  Hence the strength of the reverse
shock cannot be robustly determined for this burst.  
Although we have arbitrarily adjusted the ejecta Lorentz factor
to fit this single data point, it does not allow us to further
constrain any of the afterglow parameters, including the density.

\begin{figure*}
\includegraphics{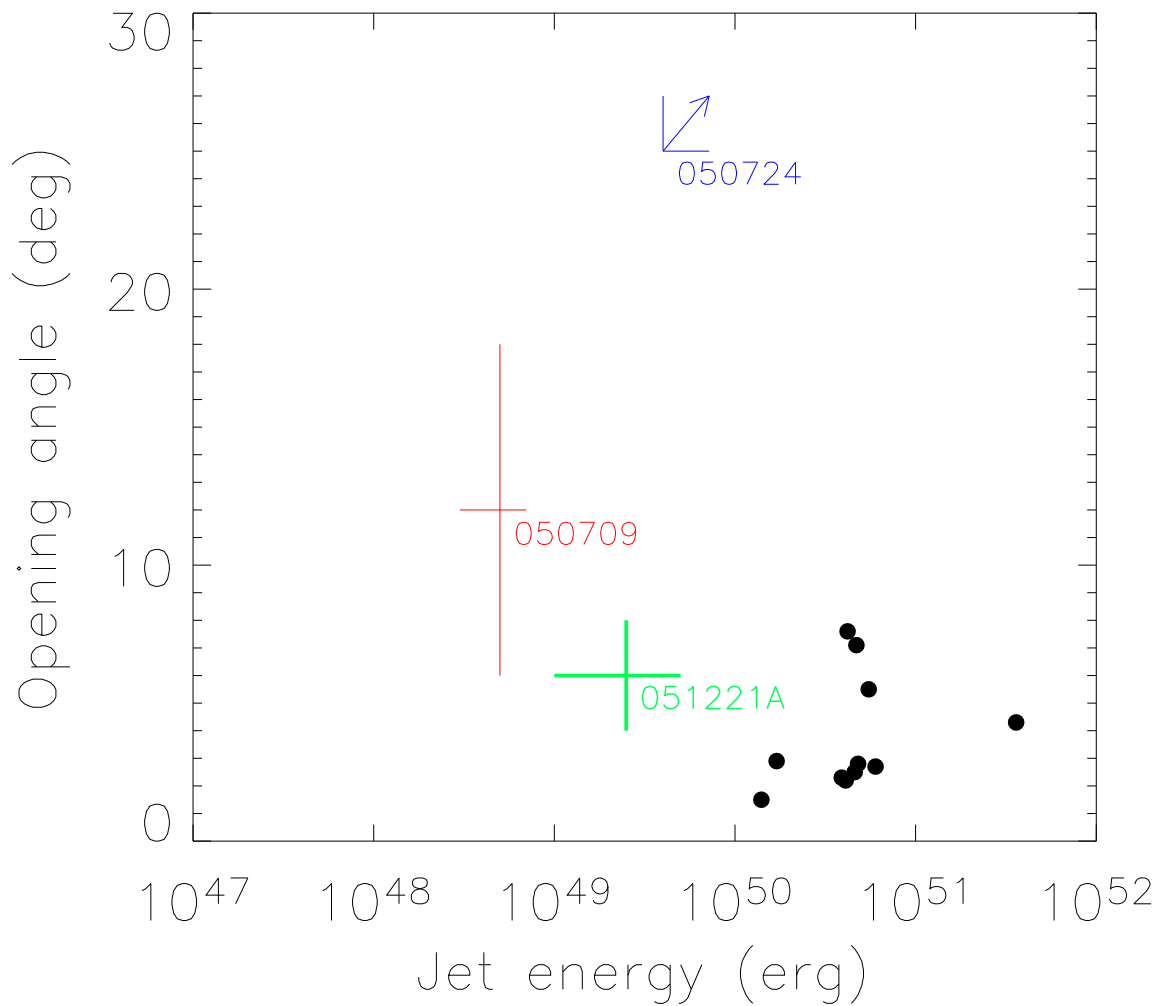}
\caption{Opening jet angle $\theta_0$ and jet energy $E_{jet}$ for 
  several long and short GRBs.  The black dots indicate values
  obtained 
for the long GRBs 980519, 990123, 990510, 991216, 000301c,
  000926, 010222, 011211, 020813, 030226 \citep[all
  from][using the same method employed here]{Panaitescu05}, and 030329 (previously unpublished result
  obtained with the same method).  
  Results for three short GRBs are
  also plotted.  For GRB\,050709 (red), we show the range of values
  allowed for the high density case obtained by \citet{Panaitescu06}.  
  For GRB\,050724 (blue) we show 
  the lower limits obtained by \citet{Grupe06}.  
  Both $\theta_0$ and
  $E_{jet}$ have lower limits, which are indicated by the horizontal
  and vertical lines with the angled arrow.    The
  range of values found here for GRB\,051221A is shown in green.
}
\label{fig:jets}
\end{figure*}

In Figure~\ref{fig:jets} we compare the initial jet angle, $\theta_0$,
and total jet energy, $E_{jet}$, for GRB\,051221A with those of two
other short bursts, GRB\,050709 and GRB\,050724, and with several long bursts.
For the long bursts, we show the best-fit parameters found by
\citet{Panaitescu05} using the same numerical modelling approach used
here.  
For GRB\,050709 we show the range of values obtained for the higher
density solution ($10^{-4}~{\rm cm}^{-3} < n < 0.1~{\rm cm}^{-3}$)
found by \citet{Panaitescu06}; a jet angle larger
than most of the long GRBs in this sample is required for this case.  
For GRB\,050724, we show the
limits obtained in Paper I using our late-time \chandra\ observation that
provides evidence against any jet break at less than 22 days
post-burst.
The jet angle we obtain for GRB\,051221A is consistent with jet angles
found for long GRBs \citep{Panaitescu05,Frail01,Bloom03}, but is
significantly lower than that of GRB\,050724,
implying a wide range in jet collimation for short bursts.  


We find that GRB\,051221A has a jet energy lower
than that of long GRBs by an order of magnitude.
The lower energy of the GRB\,051221A jet is consistent with an origin from a binary merger,
as in this case the mass of the debris torus formed during the merger is expected to be
$\sim 10$ times lower than that of the torus formed in the collapse of the core of massive stars.
Furthermore, the lower limit of circumburst densities allowed for the
short GRBs 050709 and 051221A is
compatible with these bursts occurring outside their host galaxies, as expected from
the large kick velocities that neutron stars can acquire at birth.

The flat portion of the light curve of GRB\,051221A is very similar to
flat segments seen in many long GRBs.  
In fact, this light curve looks very similar to that of GRB\,050315
\citep{Vaughan06}, with the notable exception that GRB\,051221A has no
evidence for a steeply decaying initial segment.
These flat segments have been interpreted as
being due to energy injection into the external shock
\citep{Nousek06,Zhang06,Panaitescu06b}, but this is the first time
that this behavior has been seen in a short GRB.  As discussed more
extensively by \citet{Soderberg06}, the implication is
that the external shock continues to be refreshed hours after the GRB
itself, either due to continued activity from the central engine or to
lower velocity shocks from the initial burst catching up with the decelerating blast wave.
By contrast, GRB\,050724 has a large, late bump in the X-ray light curve, interpreted as
continued central engine activity (Paper I), with the post-flare
afterglow returning to the same power-law decay as the pre-flare afterglow.  
No flares are seen in the afterglow of GRB\,051221A.

\section{Conclusions}

We find that the X-ray afterglow of GRB\,051221A has a break in the
light curve between 2.9 and 9.1 days post-burst, from a decay index of 1.2 to
$\sim 2$, that is consistent with being a jet break.
Unfortunately, only upper limits are available in the optical and
radio data following this break, so the achromaticity of the break
cannot be established; however, the break can be modelled as a jet
interacting with an external medium with a density in the range
$10^{-4} < n < 0.1$~cm$^{-3}$.  
At the low density limit we find an initial jet opening angle of
4\degrees\ and a total jet kinetic energy of $10^{49}$ ergs.  
At the higher density limit, which may be more consistent with the
small offset of the afterglow from the host galaxy and with the
indication of a modest amount of intrinsic X-ray absorption in the host, we
find a jet angle of 8\degrees\ and jet energy of $5 \times 10^{49}$ ergs. 
We obtain a much wider range of possible circumburst densities than found
by \citet{Soderberg06} and somewhat larger jet energies (by a factor
of 1.5-7), but are in agreement with their results for the remaining
jet parameters.
These results indicate that at least some short GRBs have afterglows
collimated to angles similar to jets in long GRBs, though the jet
energy can be substantially lower.  Together with the lack of a jet
break in GRB\,050724 (Paper I), this implies a wide range of jet
angles for short GRBs.  
Finally, the X-ray light curve of GRB\,051221A has a strong energy
injection signature, indicating that the energy of the external shock
is increased by a factor of two at about 0.1 days after the burst.

\acknowledgments

This research was supported by NASA contract NAS5-00136, SAO grant
GO6-7050A, and ASI grant I/R/039/04.
We acknowledge the use of data obtained through the High Energy 
Astrophysics Science Archive Research Center (HEASARC) Online Service, provided by the 
NASA/Goddard Space Flight Center.
Special thanks to Paul Plucinsky, Andrea Prestwich, and other members of the
\chandra\ X-ray Center staff who helped plan and execute the \chandra\ Target of
Opportunity observations across the holidays, and who performed the
rapid data processing needed to schedule the entire set of observations.

\clearpage

\begin{deluxetable}{lccccc}
\tablecaption{{\it Chandra}/ACIS-S Observation Log for GRB\,051221A
\label{tbl:obs_log}}
\tablewidth{0pt}
\tablehead{
\colhead{ObsID} &
\colhead{Date} &
\colhead{$\rm T_{start}$ (UT)} &
\colhead{$\rm T_{end}$ (UT)} &
\colhead{$\rm T_{elapse}$\tablenotemark{1}} &
\colhead{$\rm T_{exp}$ (ks)} 
}
\startdata
6681 & 22 Dec 2005 & 13:55:20 & 22:51:11 & 0.13 & 29.8 \\
7256 & 25 Dec 2005 & 12:17:00 & 21:13:04 & 0.38 & 29.8 \\
7257 &  5 Jan 2006 & 05:49:54 & 11:30:12 & 1.31 & 17.9 \\
7258 & 10 Jan 2006 & 02:11:17 & 09:44:07 & 1.73 & 24.6 \\
6683 & 16 Jan 2006 & 01:17:58 & 15:19:44 & 2.24 & 48.5 \\
\enddata
\tablenotetext{1}{$\rm T_{elapse} = T_{start} - T_0$ (in Ms), where
  $\rm T_0$ is the burst trigger time (01:51:16 UT on 21 December 2005).}
\end{deluxetable}

%
%

\begin{deluxetable}{lcccc}
\tablecaption{Power law spectral fits to the X-ray data of GRB\,051221A 
\label{tbl:spectral}}
\tablewidth{0pt}
\tablehead{
\colhead{} &
\colhead{$\Gamma$} &
\colhead{\NH\tablenotemark{1}} &
\colhead{$\chi^2 / \nu$} &
\colhead{Flux\tablenotemark{2}}
} 
\startdata
\swift/XRT (WT mode) & $2.2 \pm 0.4$ & $1.7^{+1.0}_{-0.9}$ &  22.5 / 15 & \tablenotemark{3} \\
\swift/XRT (PC mode)\tablenotemark{4} & $2.1 \pm 0.2$ & $1.8^{+0.7}_{-0.5}$ &  13.5 / 30 & \tablenotemark{3} \\
\chandra/ACIS-S  & $1.94^{+0.29}_{-0.19}$ & $1.4 \pm 0.9$ &  17.2 / 22 &  \\
\chandra/ACIS-S \#6681 & $1.88^{+0.16}_{-0.23}$ & $1.3^{+0.9}_{-0.8}$ &  15.4 / 18 & 1.7 \\
\chandra/ACIS-S \#7256 & $2.07^{+0.36}_{-0.33}$ & 1.4 $(frozen) $ & 1.6 / 4 & 0.46 \\
\enddata

\tablenotetext{1}{Absorption column density in units of $10^{21}$
cm$^{-2}$.  The Galactic column density along this line of sight is
$6.7 \times 10^{20}$ cm$^{-2}$ \citep{Dickey90}.}

\tablenotetext{2}{Flux units are $10^{-13}$ erg cm$^{-2}$ s$^{-1}$ for
  the 0.3--10 keV energy band.}

\tablenotetext{3}{Flux is not quoted for XRT data because it changes
  rapidly during these observation intervals (see Fig. 2 for flux vs. time).}

\tablenotetext{4}{PC mode data from T+501~s to T+65721~s, excluding
  the first portion, which is piled up.}

\end{deluxetable}

\begin{deluxetable}{cccl}
\tablecaption{X-ray light curves for GRB\,051221A 
\label{tbl:lc}}
\tablewidth{0pt}
\tablehead{
\colhead{$T_{start}$ (ks)} &
\colhead{$T_{end}$ (ks)} &
\colhead{Flux (erg cm$^{-2}$ s$^{-1}$)} &
\colhead{Instrument / Mode}
} 
\startdata
  0.092   &   0.112     &   $ (1.46 \pm 0.22) \times 10^{-10} $ &  \swift\ XRT / WT \\
  0.112   &   0.132     &   $ (1.95 \pm 0.21) \times 10^{-10} $ &  \swift\ XRT / WT \\
  0.132   &   0.152     &   $ (1.58 \pm 0.19) \times 10^{-10} $ &  \swift\ XRT / WT \\
  0.152   &   0.172     &   $ (1.25 \pm 0.17) \times 10^{-10} $ &  \swift\ XRT / WT \\
  0.172   &   0.192     &   $ (8.98 \pm 1.74) \times 10^{-11} $ &  \swift\ XRT / WT \\
  0.192   &   0.212     &   $ (1.08 \pm 0.19) \times 10^{-10} $ &  \swift\ XRT / WT \\
  0.212   &   0.252     &   $ (8.62 \pm 1.17) \times 10^{-11} $ &  \swift\ XRT / WT \\
  0.252   &   0.332     &   $ (5.47 \pm 0.84) \times 10^{-11} $ &  \swift\ XRT / WT \\
  0.301   &   0.401     &   $ (4.74 \pm 0.66) \times 10^{-11} $ &  \swift\ XRT / PC  \\
  0.401   &   0.501     &   $ (2.97 \pm 0.53) \times 10^{-11} $ &  \swift\ XRT / PC  \\
  0.501   &   0.701     &   $ (2.44 \pm 0.33) \times 10^{-11} $ &  \swift\ XRT / PC  \\
  0.701   &   0.901     &   $ (2.17 \pm 0.31) \times 10^{-11} $ &  \swift\ XRT / PC  \\
  0.901   &   1.101     &   $ (1.06 \pm 0.22) \times 10^{-11} $ &  \swift\ XRT / PC  \\
  1.101   &   1.301     &   $ (9.32 \pm 2.03) \times 10^{-12} $ &  \swift\ XRT / PC  \\
  1.301   &   1.701     &   $ (8.87 \pm 1.40) \times 10^{-12} $ &  \swift\ XRT / PC  \\
  1.701   &   2.101     &   $ (6.43 \pm 1.20) \times 10^{-12} $ &  \swift\ XRT / PC  \\
  2.101   &   2.501     &   $ (4.88 \pm 1.04) \times 10^{-12} $ &  \swift\ XRT / PC  \\
  6.101   &   7.101     &   $ (2.95 \pm 0.51) \times 10^{-12} $ &  \swift\ XRT / PC  \\
  7.101   &   8.101     &   $ (2.31 \pm 0.45) \times 10^{-12} $ &  \swift\ XRT / PC  \\
  12.10   &   13.10     &   $ (3.13 \pm 0.53) \times 10^{-12} $ &  \swift\ XRT / PC  \\
  13.10   &   14.10     &   $ (2.49 \pm 0.47) \times 10^{-12} $ &  \swift\ XRT / PC  \\
  17.40   &   18.78     &   $ (2.09 \pm 0.35) \times 10^{-12} $ &  \swift\ XRT / PC  \\
  18.78   &   19.91     &   $ (1.72 \pm 0.33) \times 10^{-12} $ &  \swift\ XRT / PC  \\
  23.27   &   31.48     &   $ (1.53 \pm 0.17) \times 10^{-12} $ &  \swift\ XRT / PC  \\
  34.97   &   43.05     &   $ (1.19 \pm 0.15) \times 10^{-12} $ &  \swift\ XRT / PC  \\
  46.99   &   54.62     &   $ (7.63 \pm 1.29) \times 10^{-13} $ &  \swift\ XRT / PC  \\
  57.77   &   65.71     &   $ (6.34 \pm 1.01) \times 10^{-13} $ &  \swift\ XRT / PC  \\
  75.13   &   77.76     &   $ (5.56 \pm 1.18) \times 10^{-13} $ &  \swift\ XRT / PC  \\
  81.55   &   93.55     &   $ (5.05 \pm 1.04) \times 10^{-13} $ &  \swift\ XRT / PC  \\
  93.55   &   105.55    &   $ (3.98 \pm 0.92) \times 10^{-13} $ &  \swift\ XRT / PC  \\
  110.5   &   135.5     &   $ (2.21 \pm 0.43) \times 10^{-13} $ &  \swift\ XRT / PC  \\
  139.9   &   164.3     &   $ (2.11 \pm 0.44) \times 10^{-13} $ &  \swift\ XRT / PC  \\
  167.9   &   199.2     &   $ (1.24 \pm 0.29) \times 10^{-13} $ &  \swift\ XRT / PC  \\
  202.7   &   233.9     &   $ (1.49 \pm 0.33) \times 10^{-13} $ &  \swift\ XRT / PC  \\
  237.4   &   268.6     &   $ (1.23 \pm 0.29) \times 10^{-13} $ &  \swift\ XRT / PC  \\
  272.1   &   303.3     &   $ (9.48 \pm 2.52) \times 10^{-14} $ &  \swift\ XRT / PC  \\
  307.3   &   337.8     &   $ (8.02 \pm 2.35) \times 10^{-14} $ &  \swift\ XRT / PC  \\
  341.5   &   424.8     &   $ (6.17 \pm 1.35) \times 10^{-14} $ &  \swift\ XRT / PC  \\
  428.3   &   511.5     &   $ (3.63 \pm 1.07) \times 10^{-14} $ &  \swift\ XRT / PC  \\
  515.1   &   597.6     &   $ (3.66 \pm 1.12) \times 10^{-14} $ &  \swift\ XRT / PC  \\
  601.9   &   943.7     &   $ (1.66 \pm 0.40) \times 10^{-14} $ &  \swift\ XRT / PC  \\
  944.1   &   1199.9    &   $ < 1.68 \times 10^{-14} $          &  \swift\ XRT / PC \\ 
\hline
  130.9   &   136.3     &   $ (1.85 \pm 0.19) \times 10^{-13} $ & \chandra\ ACIS \\
  136.3   &   141.8     &   $ (1.79 \pm 0.18) \times 10^{-13} $ & \chandra\ ACIS \\
  141.8   &   148.2     &   $ (1.56 \pm 0.16) \times 10^{-13} $ & \chandra\ ACIS \\
  148.2   &   154.8     &   $ (1.51 \pm 0.15) \times 10^{-13} $ & \chandra\ ACIS \\
  154.8   &   161.1     &   $ (1.44 \pm 0.15) \times 10^{-13} $ & \chandra\ ACIS \\
  384.1   &   393.6     &   $ (5.15 \pm 0.75) \times 10^{-14} $ & \chandra\ ACIS \\
  393.6   &   404.0     &   $ (4.69 \pm 0.68) \times 10^{-14} $ & \chandra\ ACIS \\
  404.0   &   414.3     &   $ (4.49 \pm 0.67) \times 10^{-14} $ & \chandra\ ACIS \\
  1311.9  &   1329.8    &   $ (4.64 \pm 1.65) \times 10^{-15} $ & \chandra\ ACIS \\ 
  1726.7  &   1751.3    &   $ (3.20 \pm 1.18) \times 10^{-15} $ & \chandra\ ACIS \\ 
  2247.8  &   2296.3    &   $ (1.57 \pm 0.61) \times 10^{-15} $ & \chandra\ ACIS \\ 

\enddata
\end{deluxetable}



\clearpage

\end{document}